\documentclass[pra, showkeys, showpacs]{revtex4}
\usepackage{graphicx}
\usepackage{amsmath}
\usepackage{amsfonts} 
\newcommand{\Eop}{\mathsf{E}\kern-1pt\llap{$\vert$}}
\newcommand{\Mew}[2]{\mathsf{E}\kern-1pt\llap{$\vert$}(#1;#2)}
\newcommand{\OpUnit}{\mathbb{I}}
\newcommand{\MatUnit}{1\kern-3pt 1}
\newcommand{\Trace}{\textrm{Tr}}
\newcommand{\Prob}{\textrm{Prob}}
\newcommand{\Bra}[1]{\langle{#1}|}
\newcommand{\Ket}[1]{|{#1}\rangle}
\newcommand{\BraKet}[2]{\langle #1 \vert #2 \rangle}
\newcommand{\Wop}{\hat{\mathsf{W}\kern-1pt\llap{$-$}}}
\newcommand{\WOp}[1]{\hat{\mathsf{W}\kern-1pt\llap{$-$}}(#1)}
\newcommand{\Rnumb}{\mathbb{R}}
\newcommand{\Cnumb}{\mathbb{C}}

\newcommand{\Znumb}{\mathbb{Z}}

\newcommand{\Mod}[1]{\left| #1 \right|} % absolute value of a number
\newcommand{\PEv}{\textsl{PEv}\ } % projection evolution
\newcommand{\OP}[1]{\mathcal{OP}(#1)} % set of projections for other processes
\newcommand{\BS}[1]{$\textsl{BS}_#1$} % beam splitter nr
\newcommand{\PS}{\textsl{PS}\ } % phase shifter
\newcommand{\D}[1]{$\textsl{D}_#1$} % detector nr
\newcommand{\ST}{\textbf{X}} % space--time
\newcommand{\K}{\mathcal K} % state space

\begin{document}
\title{Projection evolution  in quantum mechanics\footnote{This paper has
been  supported by the Polish Ministry of Scientific Research and Information
Technology under the  (solicited) grant No PBZ-MIN-008/P03/2003.}}
\author{A.~G\'{o}\'{z}d\'{z}}
\email{Andrzej.Gozdz@umcs.lublin.pl}
\author{M.~Pietrow}
\email{mrk@kft.umcs.lublin.pl}
\author{M.~D\c{e}bicki}
\email{mariuszd@kft.umcs.lublin.pl}
\affiliation{Institute of Physics, University of Marie
Curie--Sk\l{}odowska,\\
pl.\ M.~Sk\l{}odowskiej--Curie 1,  20--031 Lublin, Poland.}
\date{\today}
\begin{abstract}
We propose a model of  time evolution of quantum objects which unites the
unitary evolution and the measurement procedures. The model allows to treat the
time on equal footing with other dynamical variables.
\end{abstract}
\maketitle
\section{Introduction} \label{Intro}
One of the main goals of quantum theory formalism is a description of time
evolution of quantum systems.  Usually the evolution is regarded as a two step
procedure which includes reversible, unitary evolution and irreversible
"measurement" handled by the so called  projection or reduction postulate.
There are some models which come across attempts to unify both,
qualitatively different steps as for example 
\cite{Grif03,Joos85, Ghir86, Blan95, Dios98, Dick98, Bub99, Pear99, Gamb98}.
However some of the major questions are not satisfactory solved so far.

The main directions of the investigations are related to a few major
modifications and interpretations of quantum theory like  consistent histories
approaches  \cite{Grif03,Omne99,Omne92}, modal interpretations \cite{Diek98,
Bub99}, quantum jumps and collapse ideas  \cite{Ghir86, Plen98, Pear99}
and different types of models based on stochastic equations  \cite{Gisi84,
Dios98, Dios89, Gisi92, Gisi93, Loub01, Davi01}. 

The measurement process is treated differently  depending on  interpretation 
of quantum mechanics. There is no place here to discuss all the possible 
interpretations. A good, but because of  very dynamic development of the field,
not quite up-to-date  review can be found in \cite{Busc96, Mitt04}, see also  
\cite{Kamp88, Ball90}  and references therein. 

The fundamental part of  description of measurement seems to be the projection 
postulate which, from the orthodox point of view, determines the state of the
system just after measurement. However, this procedure leads to inconsistencies
with the Schr\"odinger equation. In other words, the usual description of a
process of  quantum time evolution   requires a coexistence of two 
different evolution mechanisms. This leads to additional problems like the time
duration  and dynamics of the  "wave function collapse"  \cite{Myat00,
Egus00,Gisi84}, a general problem of   physical reality before and after
measurement \cite{Ball88, Ball70, Ball92} and non--local--like behavior in case
of spatially separated quantum subsystems \cite{Eins35}.

This short introduction is, of course, far from being complete, but we can 
conclude that the problems mentioned above are still opened and  their
explanation requires probably  fundamental changes of the evolution postulates 
of quantum theory.  

The goal of this paper is to propose  an alternative  model of quantum
evolution.  In the following, we work in   a nearly standard framework of
quantum mechanics replacing only the unitary time evolution and projection
postulate by a projection evolution (\PEv) idea. It means that \emph{we reject 
the Schr\"odinger equation} as  the fundamental equation of motion. It can be
recovered as a special case of the projection evolution. 

The proposed evolution procedure  unites both the unitary evolution and the
process of measurement.  Within this idea the evolution is only a sequence of
some projections made randomly by Nature,  and with a system dependent
probability distribution. The sequence of projections is dependent on 
structure of the physical system under consideration.  The
procedure can be extended to a POVM scheme \cite{Busc96}, as well. 
 
The main difference among our proposal and other major models is that the
\PEv formalism allows to treat time on equal footing with the
spatial variables. The \emph{time is not a parameter but a dynamical variable}.
This feature makes  the model hardly comparable with the ones  in which the
physical time is the evolution parameter.

Nevertheless, some aspects of the \PEv idea are similar to well
known models. In this sense, this idea lies somewhere 'between' the decoherent
histories approach \cite{Grif03} and some modal interpretations (see pp.: 108,
179, 199, 253 in) \cite{Diek98}. In fact, in the decoherent  histories 
approach the paths of evolution are determined partially by a family of
decompositions of unity. On the other hand, the  preferred basis can define
such decomposition  in the modal theories.  These decompositions of unity
describe the physical, basic properties of  quantum systems and play a role 
the  evolution operator within the \PEv approach.

One can also find some formal similarities of the projection
evolution  to  the analysis of the quantum Zeno and anti-Zeno effects e.g., 
\cite{Bala00}. However, the basic mechanism we propose does not lead to  a
sequence of subsequent measurements only, but to random choices of the next
states of the system, where the probability distribution is a function
dependent on  previous states of the system.

\medskip
The idea of the \PEv  can be treated as a special case of more
general  concept:
\begin{enumerate}
\item[1.] The states of a quantum system are described by either density  
operators or the appropriate functionals, and the term "quantum evolution"  is
understood as a sequence of state changes ordered by the the Nature.

This idea requires introducing of an ordering parameter which orders the
subsequent changes of the state of a physical system.
This parameter, denoted by $\tau$  will be called the evolution parameter.

The above statements requires additional explanations. Usually changes are
related to the time parameter. In general, in our approach, there is no direct
relation between both, the time and state changes. 
We suggest that the natural change of the state  of a physical is the primitive
notion. The space and time are the additional structures of our Universe. They
are dynamical variables in the theory.

The standard non--relativistic quantum mechanics can be considered as
approximation in which both the evolution parameter and the physical time are in
one--to--one correspondence. In this case, the time is not a dynamical
\textbf{variable}, but a parameter which can be identified with the evolution
parameter of the \PEv.
 
Both, the evolution parameter and the physical time can be approximately related
one to another for the  physical systems  described by  the states  which  are
strongly localized in the time variable.  This can explain the successes of the
traditional approach but it leads also to  some well known paradoxes. We will
consider this type of quantum systems in the the section containing a few
examples. 

Many problems like time--energy  uncertainty principle, time of arrival, 
life--time problems, the Young type experiments for time--slits 
\cite{TimeSlits} and similar phenomena suggest that the physical time should be
the dynamical variable not a parameter. It is much more natural assumption.

In addition, the relativistic quantum mechanics, in fact,  requires that
both notions the evolution parameter and the time variable should be different.
Note, that mixing of the space and time coordinates by  the Lorentz
transformations  results in  the undetermined status of transformed
coordinates. They are neither parameters nor variables. The same one can say
even about Galilean transformations. 

We claim that the physical time should be treated on the same footing as the
other spatial coordinates. It should be also the dynamical variable which spans 
the additional dimension in the physical space. 
The physical states should determine localization of the system in space and
time according to quantum rules. 

We suppose that the evolution parameter $\tau$ is a common parameter for the
whole Universe. In this sense it parameterizes a kind of the "internal clock"
of the Universe. Because $\tau$ is  the parameter  it cannot be affected by 
physical interactions (the physical time can !) 

For simplicity we assume that $\tau$ is a real number parameter.

\item[2.] There exists a World Lottery mechanism (chooser) which chooses 
randomly the  next state of  quantum system in a way dependent on 
physical structure of this quantum system. 

The realization of the chooser is based on the so called selected collapse
projection postulate (see \cite{Stra96} and references therein). Our chooser 
is not driven by the random noise  but it is a kind of the  \emph{quantum state
dependent World Lottery}.

The evolution parameter $\tau$ is an internal parameter of the "World
Lottery"  mechanism. For every $\tau$ the Nature draws lots to get  next state
of the Universe.

It means, that the projection evolution (\PEv) can be considered as a
stochastic process (in respect to $\tau$, instead of time) determining,  at
each step of evolution, a physical state of the system represented by a
density  operator $\rho$. 

\item[3.] The probability distribution for the chooser  is determined, in some
way, by the previous states of the quantum system.

\item[4.] The evolutions of  a larger  quantum system and its subsystems have
to be consistent.

Because we assume that the evolution parameter $\tau$ is common  for  our
Universe, the Universe itself changes states as a whole. It means that the
effective evolution  for a given subsystem should be induced by the global 
evolution.

The description of the global  evolution (for the whole Universe) is in
practice  not  available, and in fact, we are able to write down only evolution
operators for larger or smaller subsystems. These effective operators have to
satisfy some scaling conditions. In the next section a proposal for scaling
condition is discussed. However, this is still open problem.
\end{enumerate}
Because we expect that such  fundamental mechanism as evolution of quantum
system should be a rather simple physical law, consistent with known results of
quantum mechanics, we propose a rather natural realization of this idea in the
next paragraph.

The approach is \emph{observer--free}. We do not need neither  to split the
physical world into \emph{observer and the physical objects} which have to be
observed  nor \emph{to introduce a mind}, which allows  to choose the state
after  observation , as is required in some approaches (collapse of states is
the most natural process in the \PEv).

The \PEv do not require also  \emph{splitting of the Universe 
into classical and quantum  worlds}. 

We do not need also  any strange assumptions about \emph{our "knowledge in
which way"} e.g., the particle passes through the slits.
 
These results are  a consequence of a single  principle postulated within  the
\PEv approach.

\section{Projection evolution} 
\label{PrEv}
Let us consider a quantum system described by  states represented by the
quantum density operators  $\rho$. In the following by $\tau$ we  denote the
evolution parameter which  orders the evolution of the Universe.
In addition, for each  $\tau$  we define a family of projectors  which gives  
an orthogonal resolution of unity i.e.,\ roughly speaking, for each $\tau$ 
they fulfill the following conditions:
\begin{eqnarray}
&\Mew{\tau}{\nu}\Mew{\tau}{\nu'}=\delta_{\nu,\nu'}\Mew{\tau}{\nu} \nonumber \\
&\sum_\nu \Mew{\tau}{\nu} = \OpUnit,
\label{ResUnit} % \eqno(1)
\end{eqnarray}
where $\OpUnit$ denotes the unit operator. \\
The operators $\Mew{\tau}{\nu}$ should represent the essential constraints of
the physical system,  responsible for its time evolution. The  Hamiltonian
plays the same role in traditional approach.  In this sense, they play a role
of the evolution operators because they  force changes of the physical system
which they describe.  
The indices $\nu$ represent here the sets of quantum numbers labelling
the projection operators (\ref{ResUnit}), uniquely.

Instead of usual unitary evolution we postulate that the new state  
$\rho(\tau;\nu)$ of the physical system, at the evolution parameter $\tau$, is 
created from the previous state $\rho(\tau-d\tau;\nu')$ by \emph{randomly
chosen, projection} $\Mew{\tau}{\nu}$. We will show a few examples how to 
construct such families of projectors, in next section.  

The projectors  $\Mew{\tau}{\nu}$ are chosen randomly according to the
following  probability distribution:
\begin{equation}
\Prob(\tau;\nu)=
\Trace \bigl[ \Mew{\tau}{\nu}\rho(\tau - d\tau;\nu')\Mew{\tau}{\nu} \bigr],
\label{(!e!2)} % \eqno(2)
\end{equation}
where $\nu$ runs over the sets of quantum numbers describing the system at the
evolution parameter $\tau$ and $\nu'$ is the set of quantum numbers which
describes the previous state.
It  means, that the Nature applies the 'projection postulate' to get a new 
state  from the previous one: 
\begin{equation}
\rho(\tau;\nu)=\frac{\Mew{\tau}{\nu}\rho(\tau- d\tau;\nu')\Mew{\tau}{\nu}}
{\Trace[\Mew{\tau}{\nu}\rho(\tau -d\tau;\nu')\Mew{\tau}{\nu}]}.
\label{quant.3} % \eqno(3)
\end{equation}
To simplify our notation let us  imagine that the projections $\Mew{\tau}{\nu}$
are constant functions of the evolution parameter on the intervals 
$<\tau_k,\tau_{k+1})$, where 
$\tau_0 < \tau_1 < \tau_2 < \do\taus \tau_n, \dots$. 
Using this notation the idea of \PEv leads to the following
recurrence equation for  states of the system under consideration:
\begin{equation}
\rho(\tau_{n+1};\nu_{n+1})=\frac{\Mew{\tau_{n+1}}{\nu_{n+1}}\rho(\tau_n;\nu_n)
\Mew{\tau_{n+1}}{\nu_{n+1}}}%
{\Trace[\Mew{\tau_{n+1}}{\nu_{n+1}}\rho(\tau_n;\nu_n)
\Mew{\tau_{n+1}}{\nu_{n+1}}]}.
\label{(!e!3)} % \eqno(4)
\end{equation}
In other words, the equation (\ref{quant.3}) describes the following mechanism:
having  the state of a physical system given by $\rho(\tau_n;\nu_n)$ at the 
evolution parameter  $\tau_n$, the next state at  $\tau_{n+1}$  is chosen 
randomly, with the probability distribution (\ref{(!e!2)}), from all possible
states (\ref{(!e!3)}), where $\nu_{n+1}$ runs over the whole range required by
(\ref{ResUnit}). 

The applied projection postulate is in the form of the so--called "selective
collapse" \cite{Stra96}. An important feature of the formula (\ref{(!e!2)}) is
a conservation of "purity" of states. If the previous state is pure, the next
one is also a pure state. This allows to omit the problem of "ensemble of
worlds" as mentioned in \cite{Pear99}. It is important to note that in this 
approach we do not need any "environment"; the same procedure should be
applied for microscopic and macroscopic systems with appropriate projection
evolution operators, related by the scaling condition mentioned in the
introduction and which can be postulated as follows: having a density operator
$\rho_W$ for larger system $W$ , it is commonly assumed that the density
operator $\rho_A$  of the subsystem $A \subset W$ can be obtained by partial
trace over the residual subsystem $A'$ i.e., $\rho_A=\Trace_{A'}(\rho_W)$. To
have consistent description  both $W$ and $A$ one needs to fulfill the
following obvious condition (scaling property):
\begin{equation}
\Trace_{A'}%
\frac{\Mew{W,\tau}{\nu}\rho_W(\tau-d\tau) \Mew{W,\tau}{\nu}}
     {\Trace[\Mew{W,\tau}{\nu}\rho_W(\tau-d\tau)\Mew{W,\tau}{\nu}]}
=
\frac{\Mew{A,\tau}{\mu}\rho_A(\tau-d\tau) \Mew{A,\tau}{\mu}}
{\Trace_A[\Mew{A,\tau}{\mu}\rho_A(\tau-d\tau)\Mew{A,\tau}{\mu}]},
\label{(!e!3a)} % \eqno(5)
\end{equation}
where $\Mew{A,\tau}{\mu}$ are required effective \PEv operators
for the subsystem $A$ with the appropriate sets of quantum numbers $\mu$
corresponding, in some way, to the quantum numbers $\nu$ of the larger system
$W$. It means that drawing of lots required by the \PEv
approach should be correlated within the connected parts of the Universe.

In principle, having the state $\rho_W$ and $\Mew{W,\tau}{\nu}$ for the
Universe the condition  (\ref{(!e!3a)}) should allow obtaining the
appropriate states and the \PEv operators for required
subsystems. Till now, we have no proof for uniqueness of solutions of the
equations  (\ref{(!e!3a)}) in respect to  $\Mew{A,\tau}{\mu}$. 
   
The general equation (\ref{(!e!3)}) suggests that the process of evolution can
follow various paths 
$\nu_0\rightarrow\nu_1\rightarrow \nu_2 \rightarrow \dots \rightarrow\nu_n$
considered as the series of projections chosen randomly at the evolution
parameters  $ \tau_0, \tau_1, \dots \tau_n$.  Then one can ask about the
conditional probability of choosing of the state $\rho(\tau_n;\nu_n)$ under
condition, that the previous states chosen by the Nature are described by 
$\nu_0\rightarrow\nu_1\rightarrow \nu_2 \rightarrow \dots \rightarrow\nu_{n-1}$.  

At this point, one needs to observe that the  \PEv  generates a
set of  "quantum  histories" consisted of products of projection operators
similar to those in  \cite{Grif03}. However, the histories are ordered in
respect to the evolution parameter, not by the physical time.  In this case  
there  is no algebra of histories because the only available "histories" are
those defined by the \PEv operator which does not allow for
logical operations on the evolution paths.  For example, in notation of
\cite{Grif03}, the negation of the {\bf compound} "history" like
$Y=F_1(\tau_1) \odot F_2(\tau_2)$ has no meaning in \PEv, because \PEv
represents a step by step stochastic process and  $1-Y$ cannot be ordered by
the evolution parameter. The same one can say about the other logical
operations on the paths.

According to the \PEv idea, the main physical interest is not in the products
of projections  only, but also in series of states which they produce. More
precisely, the system is fully described at the evolution parameter $\tau$ if
we have the state and the last projection operator responsible for generating
this state from the previous one -- a kind of history of the system is a
sequence of pairs of quantum states and the appropriate projections. 

The probability of finding a given path of evolution at the evolution parameter
$\tau$ can be calculated using equations (\ref{(!e!3)}) and (\ref{(!e!2)}) and
can be written as:
\begin{eqnarray}
\Prob(\tau=\tau_n;\nu_0,\nu_1,\nu_2,\dots,\nu_n)=&
\Trace\bigl[\Mew{\tau_n}{\nu_n}\Mew{t_{n-1}}{\nu_{n-1}} \dots 
\Mew{\tau_1}{\nu_1}\Mew{\tau_0}{\nu_0}
\rho_0 \nonumber \\
&\Mew{\tau_0}{\nu_0}\Mew{\tau_1}{\nu_1}\dots \Mew{\tau_{n-1}}{\nu_{n-1}}
\Mew{\tau_n}{\nu_n}\bigr].
\label{(!e!4)} % \eqno(6)
\end{eqnarray}
In this way we have defined the  evolution of quantum systems as a kind of a 
stochastic game played by Nature. This game is not parameterized by the physical
time but the evolution parameter.  Within this idea we treat this game as a
fundamental process -- a  Law of Nature. 

\section{Examples} \label{examples}
In this section we show three simple examples of description in terms of the
projection evolution. The most important seems to be a reproducing of the
Schr\"odinger--like evolution. It is the purpose of the next subsection in
which we show how to derive the Schr\"odinger evolution using a rather  general
method of generating operator. The other examples  concern a toy model of
Mach-Zehnder interferometer and the evolution of a particle passing through the
device. 
\subsection{The harmonic oscillator}  
Let us introduce a set of mutually commuting hermitian operators 
$\WOp{\tau;1},\WOp{\tau;2}, \dots,\WOp{\tau;N}$ which  describes the essential
properties of the system within the space--time and other degrees of freedom.
It is important to notice that, due to the spectral theorem, there exists a
common decomposition of unity for all the operators $\WOp{\tau,k}$,
$k=1,2,\dots,N$.  This resolution of unity  should give the \PEv operators. 
The operators  $\WOp{\tau,1},\WOp{\tau,2}, \dots,\WOp{\tau,N}$  will be called
the evolution generating operators.

Let us consider the 3--D harmonic oscillator, but in the non--relativistic 
space--time. For this purpose we introduce the single particle state space 
(without spin) ${\cal K}=L^2(R^4)$ in the four dimensional real space $R^4$. 
The scalar product in ${\cal K}$ is defined as:
\begin{equation}
\BraKet{\Psi_1}{\Psi_2}=\int_{\Rnumb^4} dt\, d^3\vec{x} \,
\Psi_1(t,\vec{x})^\star \Psi_2(t,\vec{x})
\label{gener:eq.1} % \eqno(7)
\end{equation}
and it contains the integration over the time. It means that the time is 
dynamical variable.

Now we define the evolution generating operators as:
\begin{eqnarray}
\Wop(\tau;1)=i\hbar\frac{\partial}{\partial t}-\hat{H} \nonumber \\
\WOp{\tau;2}=\hat{L}^2, \nonumber \\
\WOp{\tau;3}=\hat{M}, 
\label{gener:eq.2} % \eqno(8)
\end{eqnarray}
where $\hat{H}$ denotes the usual harmonic oscillator Hamiltonian, $\hat{L}^2$
is   the square of the angular momentum operator and $\hat{M}$ denotes   the
third component of the angular momentum operator.

All the operators $\Wop(\tau;k)$, $k=1,2,3$ are Hermitian and they are the 
subject of the  spectral theorem. The first operator $\Wop(\tau;1)$ has the
continuous spectrum, the second one and the third have the discrete spectrum. 
It allows to write the common spectral measure as a product of spectral 
measures of these three operators:
\begin{equation}
dM(w,l,m)= dM_{\Wop(\tau;1)}(w)M_{\Wop(\tau;2)}(l)M_{\Wop(\tau;1)}(m),
\label{spM} % \eqno(9)
\end{equation}
where $dM_{\Wop(\tau;1)}(w)$ denotes the spectral measure for $\Wop(\tau;1)$, 
$M_{\Wop(\tau;2)}(l)$, the decomposition of unity for $\hat{L}^2$ operator and
$M_{\Wop(\tau;1)}(m)$ the corresponding decomposition of unity for $\hat{M}$.

The suggested evolution operator should be of the following form:
\begin{equation}
\Mew{\tau}{w,l,m}= dM_{\Wop(\tau;1)}(w)M_{\Wop(\tau;2)}(l)M_{\Wop(\tau;1)}(m),
\label{ProjEvHO} % \eqno(10)
\end{equation}
and, in fact, it is independent of evolution parameter $\tau$. As usually, there
are known difficulties for the continuous spectrum case.  These problems can be
solved in many ways e.g.,  by making use  of the smeared out operators method
\cite{gozdz:BarRacz}.

The projection evolution operator (\ref{ProjEvHO}) leads to the  
evolution--parameter independent evolution. In this case, the first random
choice decides about the  state of the system which do not evolve further with
increasing $\tau$.
 For every $\tau$ the state chosen by the Nature is of the following form:
\begin{equation}
\Phi_{wlm}(\tau;t,\vec{x})=
\sum_n c_{nlm} e^{-\frac{i}{\hbar} (E_n(\tau)+w) t}\phi_{nlm}(\vec{x})=
e^{-\frac{i}{\hbar}w t}
\sum_n c_{nlm} e^{-\frac{i}{\hbar} E_n(\tau) t}\phi_{nlm}(\vec{x}), 
\label{gener:eq.4} % \eqno(11)
\end{equation}
where $c_{nlm}$ are numerical coefficients. The vectors $\phi_{nlm}$ are the
common  eigenvectors of the three operators:  the  harmonic oscillator
Hamiltonian $\hat{H}$, $\hat{L}^2$ and  $\hat{M}$ with the eigenvalues $E_n$,
$l(l+1)$ and   $m$, respectively. The "only" difference between the 
standard approach and \PEv is that the functions  (\ref{gener:eq.4}) are
functions of time, where time is a dynamical variable. It means that such
states are not localized in time.

As it was mentioned above, the vectors (\ref{gener:eq.4}) can be normalized to
the Dirac--delta distribution. However, in reality one can expect the states in
the form of wave  packets:
\begin{eqnarray}
\int_{\Rnumb} && dw\, a(w) \Phi_{wlm}(\tau;t,\vec{x})= \nonumber \\
&&\int_{\Rnumb} dw\, a(w) \sum_n c_{nlm} e^{-\frac{i}{\hbar} 
(E_n(\tau)+w) t}\phi_{nlm}(\vec{x})=
\int_{\Rnumb} dw\ a(w) e^{-\frac{i}{\hbar}w t}
\sum_k c_{nlm} e^{-\frac{i}{\hbar} E_n(\tau) t}\phi_{nlm}(\vec{x}). 
\label{wavePackT} % \eqno(12)
\end{eqnarray}
For example, the measurement of position in time (it is a special kind of
interaction within the space--time) leads, by definition, to the 
localization in time. For $a(w)=e^{+\frac{i}{\hbar}w t'}$ the states
(\ref{wavePackT})  are "perfectly" localized in time $t'$:
\begin{equation}
\tilde{\Phi}_{lm}(\tau;t,\vec{x})=
\delta(t-t')
\sum_k c_{nlm} e^{-\frac{i}{\hbar} E_n(\tau) t}\phi_{nlm}(\vec{x}). 
\label{LWavePackT} % \eqno(13)
\end{equation}
The states (\ref{LWavePackT}) correspond exactly to the traditional
Schr\"odinger type solutions for the harmonic oscillator. However, one needs to
realize that within the \PEv these solutions represent the states which are
localized in the fixed time $t'$. It means, that one needs to have some
additional  interactions to "shift" these states in time. This is what we are
doing  in the traditional unitary evolution approach.  Within the standard
quantum mechanics  one assumes  implicitly, that there exists a "time machine"
which evolves the  state perfectly localized in time  to later moments. This
mechanics describes the state vectors projected onto the subsequent moments of 
time loosing possible non--local time correlations.

On the other hand, the state  (\ref{wavePackT}) describes a full time behavior
of the system as a single vector which allows to calculate typical statistical
characteristics like averages, variations and other statistical moments. 
For example,  the average time position of the harmonic oscillator described by
the wave packets of the form $\ref{wavePackT}$ can be  obtained in the usual 
way by the integration:
\begin{equation}
<t>=\int_{\Rnumb}dt\int_{\Rnumb}dx\int_{\Rnumb} dw' \int_{\Rnumb} dw\,
a(w')^\star a(w) \Phi_{w'lm}(\tau;t,\vec{x})^\star\, t\,
\Phi_{wlm}(\tau;t,\vec{x})
\label{AverTHO} % \eqno(14)
\end{equation}
Obviously, this example  can be generalized to any arbitrary system which one
can describe in the traditional way.

\subsection{Beam splitter -- toy models}
In this paragraph we consider a very simplified  Mach-Zehnder interferometer.
The aim of the considerations is to show the temporal behavior of the
interferometer in two models.
 
The first model does not correspond to the Schr\"odinger type motion because
the beam splitters localize the particle.  The second model is based on
more traditional thinking about the beam splitters  as some unitary devices. In
both cases we get the similar results which differs mainly in temporal part of
the description.

The main disadvantage of both models is that they do not show, in general 
possible, spreading in time of  states of the system. Because of this, within
these toy models, the time of arrival of a particle to the detector is  a sharp
number, without random distribution. However, this type of models can be
analyzed on the elementary level and they are able to show some interesting
features of the formalism.   

\begin{figure}[ht]
\begin{center}
\includegraphics[scale=0.8]{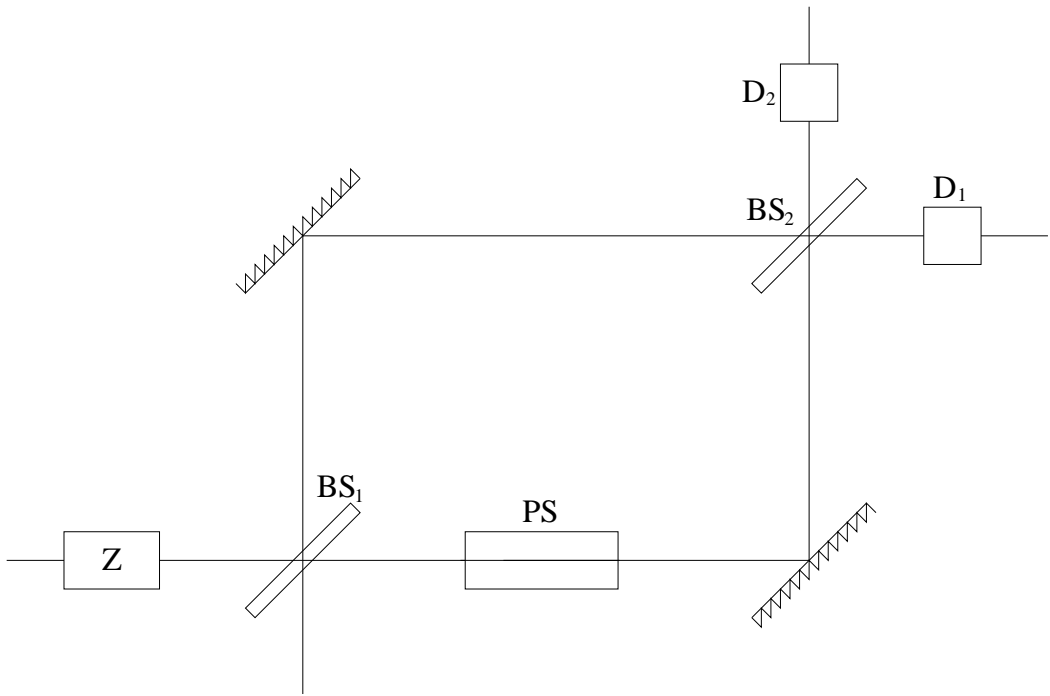}
\caption{Mach-Zehnder interferometer}
\label{fig1} % Rys.1
\end{center}
\end{figure}

\begin{figure}[ht]
\begin{center}
\includegraphics[scale=0.8]{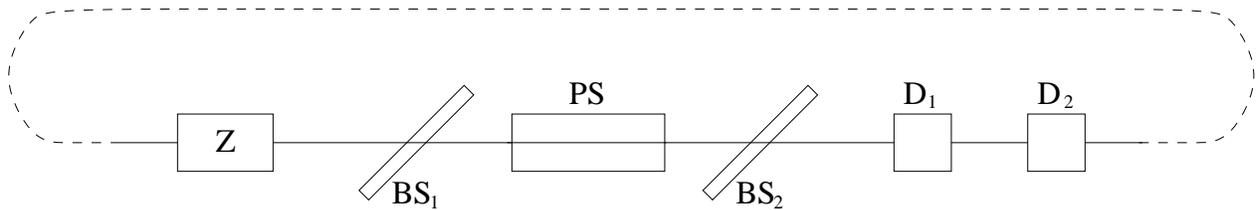}
\caption{One--dimensional Mach--Zehnder interferometer}
\label{fig2} % Rys.2
\end{center}
\end{figure}

\leftline{Toy Model I: Beam splitter which localizes events.}
\medskip
The Mach-Zehnder interferometer shown in \ref{fig1} allows to show the quantum
interference.
The particles are produced in the source \textsl{Z}.
The first beam splitter \BS{1} splits an incoming beam  into two separate
channels.
Next, there is a phase shifter \PS and the second beam splitter \BS{2}. \D{1}
and \D{2} denote the detectors which detect the particle in the first or the
second  output channel. 
However, for our purpose we consider a simplified version of the
Mach-Zehnder interferometer.
It is shown in the figure \ref{fig2}
It is able to simulate, to some extend, the Mach--Zehnder interferometer shown
in the figure \ref{fig1}  for example, it happens for  the 
$\frac{1}{2}$--spin  particles or even polarized photons.

In the following we are using the toy model (like in  sect. 2.5
of \cite{Grif03}) in which the space is discrete.
However, in our approach the time is also a dynamical variable and we have to
use not one--dimensional  space but two--dimensional  space--time for which
both the time $t$ and the coordinate $x$ takes on only  a finite numbers of
integer values:
\begin{eqnarray}
-T_a \leq & t &  \leq T_b \nonumber \\
-X_a \leq & x & \leq X_b
\label{STCoordinates} % \eqno(15) 
\end{eqnarray}
Like in \cite{Grif03} we assume the "periodic boundary conditions" for both
the time and the space coordinates, so that last sites for $t$ and $x$ are
adjacent to the first sites. In this way, our space--time \ST=$\textrm{T} 
\times \textrm{X}$ is the  Cartesian product of two  discrete "circles". First
one representing the time T consists of  $T_a+T_b+1$  points and the second one
X contains $X_a+X_b+1$ space coordinate points, respectively. In addition, we
assume an extra two--valued variable  $\zeta=a,b$ which determines two possible
channels in which each space--time point can be located (see Ch.~12 of
\cite{Grif03}). 

The space $\K$ of quantum states  consists of  complex functions:
\begin{equation}
\psi: \ST \times \{a,b\} \rightarrow \Cnumb.
\label{StateFun} % \eqno(16) 
\end{equation}
The scalar product in the state space $\K$ is:
\begin{equation}
\BraKet{\psi_1}{\psi_2}= \sum_{t',x',\zeta} \psi_1(t',x',\zeta)^\star 
                                           \psi_2(t',x',\zeta). 
\label{ScalarProd} % \eqno(17) 
\end{equation}
Again, it is very important to keep in mind, that time  is not a parameter but
a dynamical variable. It does not enumerate the subsequent events and doesn't
describe  the evolution of the system in this sense. More, it is even possible
that two subsequent, ordered by the evolution parameter $\tau$ , events   occur
in the inverse order in time. In this way we can have so called backward 
causation required for quantum world by some authors \cite{Wharton}. 

To construct the projection evolution operator we have to
analyze the physical structure (in required approximation) of devices shown in
the \ref{fig2}. 

Let us assume that the source \textsl{Z} produces the particles localized in
both the time $t$ and the position $x_Z$ i.e., in the states
$\Ket{t,x_Z,\sigma}$ which are defined  as the  "eigenfunctions" of  time,
position and  channel operators. Other words, the states $\Ket{t,x,\sigma}$ 
represent  the functions localizing the particle in  point $(t,x)$ of the
space--time and in  the channel $\sigma$:
\begin{equation}
\Ket{t,x,\sigma} \to \theta_{tx\sigma}(t',x',\zeta)= 
\delta_{tt'}\xi_{x\sigma}(x',\zeta)=
    \delta_{tt'} \delta_{xx'} \delta_{\sigma\zeta}. 
\label{LocFun} % \eqno(18) 
\end{equation}
Note that $t,x$ and $\sigma=a,b$ are here the quantum numbers but $t',x'$ and
$\zeta$ denote the corresponding variables.

Starting from the source, the particle is  created in one of the possible  states
$\Ket{t,x_Z,\sigma=a}$ i.e., at a randomly chosen time $t$, at fixed place
$x_Z$ and  in the $a$ channel. There are also possible other, less interesting
processes which are represented here by a set of projection operators 
denoted  by $\OP{Z;\nu}$ (other processes).  Together with the creation
of particle they complete a resolution of unity. At this moment we do not need 
to analyze them more carefully.

The assumptions  allow to write the following projection evolution operator 
for the source:   
\begin{equation}
\Eop_Z(\nu) = 
\begin{cases}
\Ket{t,x_Z,a}\Bra{t,x_Z,a} & \text{for}\ \nu=t\in \textrm{T} \\
\OP{Z;\nu} & \text{otherwise} 
\end{cases}
\label{PEvZ} % \eqno(19)
\end{equation}
Note that the operator is independent of $\tau$.

As the next step we have to construct the toy model of free evolution. This
type of evolution is required to describe behavior of the particle outside of
devices i.e., in the free space--time. 

In our toy model (I) we assume  the free evolution of projective type but which
resembles  the Schr\"odinger's type of evolution. For this purpose we define
the unitary transformation $S$ (see the toy models in \cite{Grif03}):
\begin{equation}
\hat{S}\Ket{x,\sigma} = \Ket{x+1,\sigma},
\label{STransform} % \eqno(20)
\end{equation}
where the ket $\Ket{x,\sigma}$ represents the function
$\xi_{x\sigma}(x',\zeta)$, see (\ref{LocFun}). The operator $\hat{S}$ is the
unitary operator within the space spanned by the kets $\Ket{x,\sigma}$ i.e.,
spanned by the functions  $\xi_{x\sigma}(x',\zeta)$. 

The last property allows to define the set of vectors which allow to define
the  Schr\"odinger's type of evolution. These vectors will be denoted by:
\begin{equation}
\Ket{\phi_{l,x,\sigma}} \rightarrow
\phi_{l,x,\sigma}(t',x',\zeta)=\chi_l(t')\hat{S}(t')\xi_{x,\sigma}(x',\zeta),
\label{PhiBase} % \eqno(21)
\end{equation}
where $\hat{S}(t')=\hat{S}^{t'}$ is the local (it is a function of time!) 
unitary operator in (\ref{STransform}). We assume that the set of functions
$\chi_l$, where $l$ is  an integer, give an orthonormal basis in the
space of all functions dependent only on time:
\begin{equation}
\BraKet{\chi_{l'}}{\chi_{l}}=
\sum_{t'} \chi_{l'}(t')^\star \chi_{l}(t')=\delta_{l'l}. 
\label{TBasis} % \eqno(22)
\end{equation}
The functions $\chi_l$ determine the main part of the amplitude describing the
fact  that the particle can be localized in time (see remarks below the eq.
(\ref{PrPsi})).

It implies that the functions $\phi_{l,x,\sigma}$  form an orthonormal basis in
the space of states $\K$. 

Because these vectors are independent of the evolution parameter,  the
projection evolution operator for free evolution can be considered as a single
random choice (like in the first example): 
\begin{equation}
\Eop_F(\nu) = 
\sum_{x,\sigma} \Ket{\phi_{\nu,x,\sigma}}\Bra{\phi_{\nu,x,\sigma}},\ 
\text{where } \nu \in \Znumb.
\label{PEvFree} % \eqno(23)		    
\end{equation}
Observation: The action of the operator (\ref{PEvFree}) on a vector
state $\psi$ gives a Schr\"odinger--type function:
\begin{eqnarray} 
\Eop_F(\nu)\psi(t',x',\zeta) &=&
 \sum_{x,\sigma} \BraKet{\phi_{\nu,x,\sigma}}{\psi}
 \phi_{\nu,x,\sigma}(t',x',\zeta)  \nonumber \\
&=& \sum_{x,\sigma} \BraKet{\phi_{\nu,x,\sigma}}{\psi}
\chi_{\nu}(t') \xi_{x+t',\sigma}(x',\zeta).
\label{PrPsi} % \eqno(24)		    
\end{eqnarray}
The Nature chooses by chance (with the distribution determined by the last
state)  the actual state labelled by $\nu$, which characterizes mainly the 
potential localization of the particle in the physical time. In this model,
time and space positions  are treated on the same  footing. It means that there
are possible situation when particle cannot be found in some regions of time.
However, this situation cannot be interpreted in the usual way that such
particle does not exist. Existence or non--existence of the particle is 
determined here  by the total state considered in the full space--time.   

Next device in our system is the beam splitter which can be thought as a device 
with two input channels and two output channels. In this model  we assume that
the beam splitters are able to localize particles in the space--time. It means
that they are not the unitary--type devices.

To construct the appropriate operators, let us define a family of vectors 
which allows to "send" a particle into both channels simultaneously:
\begin{eqnarray}
\Ket{bs_1;t\ x_{BS}\ \varepsilon} = &
\cos\varepsilon\, \Ket{t\ x_{BS}\ a} +  
\sin\varepsilon\, \left[\frac{1}{\sqrt{2}} 
\left(\Ket{t,x_{BS}+1,b} + \Ket{t,x_{BS}+1,a}\right) \right] \nonumber \\
\Ket{bs_2;t\ x_{BS}\ \varepsilon} = &
- \sin\varepsilon\, \left[\frac{1}{\sqrt{2}} 
\left(-\Ket{t,x_{BS}+1,b} + \Ket{t,x_{BS}+1,a}\right) \right]
+ \cos\varepsilon\, \Ket{t\ x_{BS}\ b} \nonumber \\ 
\Ket{bs_3;t\ x_{BS}\ \varepsilon} = &
- \sin\varepsilon\, \Ket{t\ x_{BS}\ a}  
+ \cos\varepsilon\,\left[\frac{1}{\sqrt{2}} 
\left(\Ket{t,x_{BS}+1,b} + \Ket{t,x_{BS}+1,a}\right) \right] \nonumber  \\
\Ket{bs_4;t\ x_{BS}\ \varepsilon} = &
\cos\varepsilon\, \left[\frac{1}{\sqrt{2}} 
\left(- \Ket{t,x_{BS}+1,b} + \Ket{t,x_{BS}+1,a}\right) \right]
+ \sin\varepsilon\, \Ket{t\ x_{BS}\ b} 
\label{BSVect}  % \eqno(25) 
\end{eqnarray}
As the beam splitter has two input (entrance) channels, its projection evolution
operator should  take into account these two possibilities and can be written
as the resolution of unity as follows:
\begin{equation}
\Eop_{BSn}(\tau; \nu) = 
\begin{cases}
\Ket{bs_k;t,x_{BSn},\varepsilon} \Bra{bs_k;t,x_{BSn},\varepsilon} +
\Ket{bs_{k+1};t,x_{BSn},\varepsilon} \Bra{bs_{k+1},t,x_{BSn},\varepsilon} 
& \text{for}\ \nu=(k,t),\ k=1,3; t \in \textrm{T};\\ % t nie jest w Z !
\OP{BSn,\tau;\nu} \quad  \text{otherwise} &
\end{cases}
\label{PEvBS}  % \eqno(26) 
\end{equation}
where $BSn=BS1,BS2$ denotes the first or the second beam splitter.  The beam
splitters are placed in different positions $x_{BS1}$ and $x_{BS2}$. 

The parameter $\varepsilon$ is changing continuously from $0$ to
$\frac{\pi}{2}$, while the particle is inside the device. The parameter
$\varepsilon=\varepsilon(\tau)$ is a monotonic, increasing function of the
evolution parameter. More detailed structure of the function
$\varepsilon=\varepsilon(\tau)$ is not needed.

The continuous series of projections, while $\varepsilon$ is changing  from $0$
to $\frac{\pi}{2}$, lead to the following effective  projection evolution
operator:
\begin{equation}
\Eop^{eff}_{BSn}(\nu) = 
\begin{cases}
\frac{1}{\sqrt{2}} 
\left(\Ket{t,x_{BS}+1,b} + \Ket{t,x_{BS}+1,a}\right)\Bra{t,x_{BS},a}  \\
-  \frac{1}{\sqrt{2}} 
\left(-\Ket{t,x_{BS}+1,b} + \Ket{t,x_{BS}+1,a}\right)\Bra{t,x_{BS},b} 
& \text{for}\  \nu=(1,t), t \in \textrm{T}\\
-\frac{1}{\sqrt{2}} 
\Ket{t,x_{BS},a} \left(\Bra{t,x_{BS}+1,b} + \Bra{t,x_{BS}+1,a}\right) \\
+ \frac{1}{\sqrt{2}} 
\Ket{t,x_{BS},b} \left(-\Bra{t,x_{BS}+1,b} + \Bra{t,x_{BS}+1,a}\right)
& \text{for}\  \nu=(3,t), t \in \textrm{T},
\end{cases}
\label{PEvBSEffect}  % \eqno(27)
\end{equation}
where $n=1,2$ denotes the first or the second beam splitter. As the label for
this effective step of evolution one can choose any evolution parameter
$\tau_{PS}$   chosen  from the interval of the evolution parameter required to
perform the evolution of the particle  within the phase shifter.

In case of the first beam splitter each of the input channels $a$ and $b$ 
is transformed into a  superposition of both output channels $a$ and $b$.  Next
the particle passes  the phase shifter described below and there is the  second
beam splitter  where two input channels $a$ and $b$ are again transformed into
the  proper superposition of two output channels $a$ and $b$.

The phase shifter in the one dimensional model as presented in Fig.\ref{fig2} 
acts on two channels $a$ and $b$ simultaneously. To construct its evolution
operator one can use the same idea as in the construction of (\ref{PEvBS}) and
(\ref{PEvBSEffect}). For this purpose we define a family of the following 
vectors:
\begin{eqnarray}
\Ket{ps_1;t,x_{PS},\alpha,\varepsilon} &=&
\cos\varepsilon\, \Ket{t,x_{PS},\alpha}
+ \sin\varepsilon\, e^{i\phi_\alpha}\, \Ket{t+1,x_{PS},\alpha}  \nonumber \\
\Ket{ps_2;t,x_{PS},\alpha,\varepsilon} &=&
-\sin\varepsilon\, \Ket{t,x_{PS},\alpha}
+ \cos\varepsilon\, e^{i\phi_\alpha}\, \Ket{t+1,x_{PS},\alpha},   
\label{PSVect}  % \eqno(28) 
\end{eqnarray}
where $\alpha=a,b$ labels both channels and $x_{PS}$ denotes the position of
the phase shifter.

The evolution operator for \PS is presented here by the continuous  family of 
the following projections:
\begin{equation}
\Eop_{PS}(\tau; \nu) = 
\begin{cases}
\sum_{\alpha}
\Ket{ps_k;t,x_{PS},\alpha,\varepsilon}\Bra{ps_k;t,x_{PS},\alpha,\varepsilon}
& \text{for}\ \nu=(k,t),\ k=1,2, t \in \textrm{T}\\ 
\OP{PS,\tau; \nu} \quad  \text{otherwise} &
\end{cases}
\label{PEvPS}  % \eqno(29) 
\end{equation}
where the parameter $\varepsilon$ is changing continuously from $0$ to
$\frac{\pi}{2}$, while the particle is inside the device. The parameter
$\varepsilon=\varepsilon(\tau)$ is a monotonic, increasing function of the
evolution parameter.

The proper effective projection evolution operator for the phase shifter can
be created in the same way as in case of the beam splitter:
\begin{equation}
\Eop^{eff}_{PS}(\nu) = 
\begin{cases}
 e^{i\phi_a}  \Ket{t+1,x_{PS},a}\Bra{t,x_{PS},a}
+e^{i\phi_b}  \Ket{t+1,x_{PS},b}\Bra{t,x_{PS},b}
& \text{for}\  \nu=(1,t), t \in \textrm{T} \\
-e^{-i\phi_a} \Ket{t,x_{PS},a}\Bra{t+1,x_{PS},a}
-e^{-i\phi_b} \Ket{t,x_{PS},b}\Bra{t+1,x_{PS},b}
& \text{for}\ \nu=(2,t), t \in \textrm{T}.
\end{cases}
\label{PEvPSEff}  % \eqno(30)
\end{equation}
As the label for this effective step of evolution one can choose any evolution
parameter $\tau_{PS}$   chosen  from the interval of the evolution parameters
required to perform the evolution of the particle  within the phase shifter.

The detectors are the last devices we have to describe. We assume that the first
detector is placed in $x_{D_a}$ and it is able to detect particle in the 
channel $a$ and the second detector placed in $x_{D_b}$ detects the particle in
the channel $b$. We assume also that both detectors localize particle in the
space and time. These assumptions determine uniquely the form of the projection
evolution operator for detectors:
\begin{equation}
\Eop_D(\nu)=
\begin{cases}
\Ket{t,x_{D_a},a}\Bra{t,x_{D_a},a} 
& \text{for}\ \nu=(t,a),\  t  \in \textrm{T} \\
\Ket{t,x_{D_b},b}\Bra{t,x_{D_b},b} 
& \text{for}\ \nu=(t,b),\  t  \in \textrm{T} \\
\OP{D;\nu} \quad  \text{otherwise}. & 
\end{cases}
\label{PEvD}  % \eqno(31)
\end{equation}
Of course, the process of evolution is described by one evolution operator 
collected from all of the above. They describe the subsequent events enumerated
by the evolution parameter $\tau$:
\begin{equation}
\Mew{\tau}{\nu} = 
\begin{cases}
\Eop_Z(\nu) & \text{for}\    \tau=0, \\
\Eop_F(\nu)      & \text{for}\                     0 < \tau < \tau_{BS1} \\
\Eop_{BS1}(\tau; \nu) & \text{for}\    \tau_{BS1} \leq \tau < \tau_{BS1'} \\
\Eop_F(\nu)   & \text{for}\           \tau_{BS1'} \leq \tau < \tau_{PS}   \\
\Eop_{PS}(\tau; \nu) & \text{for}\ \    \tau_{PS} \leq \tau < \tau_{PS'}  \\ 
\Eop_F(\nu)   & \text{for}\            \tau_{PS'} \leq \tau < \tau_{BS2} \\
\Eop_{BS2}(\tau; \nu)  & \text{for}\   \tau_{BS2} \leq \tau < \tau_{BS2'} \\
\Eop_F(\nu)   & \text{for}\           \tau_{BS2'} \leq \tau < \tau_{D}  \\
\Eop_D(\nu) & \text{for}\    \tau=\tau_{D}.
\end{cases}
\label{PEvTotal}  % \eqno(32)
\end{equation}
The first line describes the emission of  particle in the source \textsl{Z}. The
second line gives the free motion between the source and the first beam
splitter. It can be labelled by any evolution parameter $\tau_{F1} \in 
(0,\tau_{BS1})$ because the projection of the projection is again the same
projection. The third line describes the beam splitter and it can be replaced by
the effective operator (\ref{PEvBSEffect}) assuming the evolution parameter
equal e.g., to $\tau_{BS1}$.  In fact, one can choose any evolution parameter 
from the interval  $[\tau_{BS1},\tau_{BS1'})$. Next, we have free motion between
the beam splitter and  the phase shifter. For the phase shifter, described by
the next line, one can also introduces the effective operator (\ref{PEvPSEff}) 
and the evolution parameter $\tau_{PS}$. Next two lines describe the second beam
splitter which can be interpreted in terms of the effective operators at the
evolution parameter $\tau_{BS2}$ and two detectors which detect, or not, the
particle at the evolution parameter $\tau_D$. 

In the following we are not interested in full analysis of all possibilities
given by the projection evolution operator (\ref{PEvTotal}). We want to find
the states and  the appropriate probabilities of the particle detected   at the
evolution parameter $\tau_D$.

According to general rules of \PEv the interesting states can be calculated as:
\begin{eqnarray}
&&\rho(\tau_D;\nu_Z,\nu_{F1},\nu_{BS1},\nu_{F2},\nu_{PS},\nu_{F3},
\nu_{BS2},\nu_{F4},\nu_D) = \nonumber \\
&&\ \frac{
\Eop_D(\nu_D)\Eop_F(\nu_{F4})\Eop^{eff}_{BS2}(\nu_{BS2})
\Eop_F(\nu_{F3})\Eop^{eff}_{PS}(\nu_{PS})\Eop_F(\nu_{F2})
\Eop^{eff}_{BS1}(\nu_{BS1})\Eop_F(\nu_{F1})\Eop_Z(\nu_Z)
\rho_0
\Eop_Z(\nu_Z)}{
\Trace\left[  
\Eop_D(\nu_D)\Eop_F(\nu_{F4})\Eop^{eff}_{BS2}(\nu_{BS2})
\Eop_F(\nu_{F3})\Eop^{eff}_{PS}(\nu_{PS})\Eop_F(\nu_{F2})
\Eop^{eff}_{BS1}(\nu_{BS1})\Eop_F(\nu_{F1})\Eop_Z(\nu_Z)
\rho_0
\Eop_Z(\nu_Z) \right.}   \nonumber \\
&&\quad\quad \frac{
\Eop_F(\nu_{F1})\Eop^{eff\dagger}_{BS1}(\nu_{BS1})
\Eop_F(\nu_{F2})\Eop^{eff\dagger}_{PS}(\nu_{PS})\Eop_F(\nu_{F3})
\Eop^{eff\dagger}_{BS2}(\nu_{BS2})\Eop_F(\nu_{F4})\Eop_D(\nu_D)}{
\left. \Eop_F(\nu_{F1})\Eop^{eff\dagger}_{BS1}(\nu_{BS1})
\Eop_F(\nu_{F2})\Eop^{eff\dagger}_{PS}(\nu_{PS})\Eop_F(\nu_{F3})
\Eop^{eff\dagger}_{BS2}(\nu_{BS2})\Eop_F(\nu_{F4})\Eop_D(\nu_D)
\right]} 
\label{StateRho}  % \eqno(33) 
\end{eqnarray}
The most interesting path of the evolution is given by the following
sequence:\\ $h=(\nu_Z=t_Z,\nu_{F1}, \nu_{BS1}=(1,t_{BS1}), \nu_{F2},
\nu_{PS}=(1,t_{PS}),  \nu_{F3}, \nu_{BS2}=(1,t_{BS2}),
\nu_{F4},\nu_D=(t_D,\sigma_D))$. \\ 
Obviously, the whole path is chosen randomly in step by step procedure, as
described in \PEv. 

Because the final state is a pure state, it can be immediately written as:
\begin{equation}
\rho(\tau_D;\nu_Z,\nu_{F1},\nu_{BS1},\nu_{F2},\nu_{PS},\nu_{F3},
\nu_{BS2},\nu_{F4},\nu_D) =
\Ket{t_D,x_{D_{\sigma_D}},\sigma_D}\Bra{t_D,x_{D_{\sigma_D}},\sigma_D}. 
\label{FinalState}  % \eqno(33)
\end{equation}
The distribution of times $t_D$ of detection of the particle in the channel 
$\sigma_D$ can be calculated as the conditional probability:
\begin{eqnarray}
&&\Prob(\tau_D;\nu_Z,\nu_{F1},\nu_{BS1},\nu_{F2},\nu_{PS},\nu_{F3},
\nu_{BS2},\nu_{F4},\nu_D) = \nonumber \\
&&\  \Trace\left[  
\Eop_D(\nu_D)\Eop_F(\nu_{F4})\Eop^{eff}_{BS2}(\nu_{BS2})
\Eop_F(\nu_{F3})\Eop^{eff}_{PS}(\nu_{PS})\Eop_F(\nu_{F2})
\Eop^{eff}_{BS1}(\nu_{BS1})\Eop_F(\nu_{F1})\Eop_Z(\nu_Z)
\rho_0
\Eop_Z(\nu_Z) \right.   \nonumber \\
&&\quad\quad 
\left. \Eop_F(\nu_{F1})\Eop^{eff\dagger}_{BS1}(\nu_{BS1})
\Eop_F(\nu_{F2})\Eop^{eff\dagger}_{PS}(\nu_{PS})\Eop_F(\nu_{F3})
\Eop^{eff\dagger}_{BS2}(\nu_{BS2})\Eop_F(\nu_{F4})\Eop_D(\nu_D)
\right]. 
\label{ProbDet}  % \eqno(33) 
\end{eqnarray}
The conditional probability (\ref{ProbDet}) gives the probability of detection
of the particle in the state labelled by $\nu_D$ under the conditions that the
particle is created in $\nu_Z$ an then evolves through
$\nu_{F1},\nu_{BS1},\nu_{F2},\nu_{PS},\nu_{F3}, \nu_{BS2}$ and $\nu_{F4}$.

Now assuming the initial state is the pure state denoted by: 
\begin{equation}
\rho_0 = \Ket{\psi_0}\Bra{\psi_0},
\label{InitState}  % \eqno(34)
\end{equation}
the particle will be created from the source with the probability 
$\left|\BraKet{t_Z,x_Z,a}{\psi_0}\right|^2$  and after the creation, the state
of the particle is
\begin{equation}
\Ket{\psi_Z} = \Ket{t_Z,x_Z,a}.
\label{SourceState}  % \eqno(35)
\end{equation}
Let us calculate a  scalar product  which will be useful in further
considerations:
\begin{equation}
\BraKet{\phi_{\nu,x,\sigma}}{t_Z,x_Z,a} = 
\sum_{t',x',\zeta} \chi_{\nu}(t')^* \xi_{x+t',\sigma}(x',\zeta)
\delta_{t_Z t'}\xi_{x_Z a}(x',\zeta) = 
\chi_{\nu}(t_Z)^* \xi_{x+t_Z,\sigma}(x_Z,a).
\label{UseScPr}  % \eqno(36)
\end{equation}
After the particle is created from the source,  it undergoes a free evolution.
Making use of (\ref{PrPsi})  and (\ref{UseScPr}) one can write the state of the
particle as:
\begin{eqnarray}
\Ket{\psi_{F1}} = \mathcal{N}\ \Eop_F(\nu_{F1})\Ket{t_Z,x_Z,a} = 
\mathcal{N} \sum_{x,\sigma} \Ket{\phi_{\nu_{F1},x,\sigma}}
\BraKet{\phi_{\nu_{F1},x,\sigma}}{t_Z,x_Z,a}
%\chi_{\nu}(t') \xi_{x+t',\sigma}(x',\zeta) = \\
 = \Ket{\phi_{\nu_{F1},x_Z-t_Z,a}}. 
\label{StateF1} % \eqno(37)
\end{eqnarray}
The probability that the particle is found in such a state, under the condition
that it was is in the state (\ref{SourceState}), is equal
$\Mod{\chi_{\nu_{F1}}(t_Z)}^2$. The coefficient $\mathcal{N}$ is the
normalization factor which is not needed for further considerations.

Next we can derive a state of the particle in the next step of the evolution, 
i.e. in the beam splitter BS1. Once again we calculate explicitly the scalar 
product
\begin{eqnarray}
\BraKet{t_{BS1},x_{BS1},\sigma}{\psi_{F1}} = 
\sum_{t',x',\zeta} 
\delta_{t_{BS1} t'}\xi_{x_{BS1} \sigma}(x',\zeta)
\chi_{\nu_{F1}}(t') \xi_{x_Z+(t'-t_Z),a}(x',\zeta) = \nonumber\\
\chi_{\nu_{F1}}(t_{BS1}) \xi_{x_Z+(t_{BS1}-t_Z),\sigma}(x_{BS1},a).
\label{ScPrBS1F1} % \eqno(38) 
\end{eqnarray}
This  scalar product is not equal zero, for the time $t=t_{BS1}$ only, i.e. 
it must satisfy the condition \\
$\Mod{\chi_{\nu_{F1}}(t_{BS1})}^2 \delta_{x_Z+(t_{BS1}-t_Z),x_{BS1}} \not= 0$.
It allows to write the state of the particle (for this step of the evolution)
as: 
\begin{equation}
\Ket{\psi_{BS1}} = \frac{1}{\sqrt{2}} 
\left(\Ket{t_{BS1},x_{BS1}+1,b} + \Ket{t_{BS1},x_{BS1}+1,a}\right)
\label{StateBS1} % \eqno(39)
\end{equation}
In the similar way one can find the state in the phase shifter PS  after a free
evolution F2. It is chosen by the Nature, with the  conditional probability
$
\Mod{\chi_{\nu_{F2}}(t_{BS1})}^2
\Mod{\chi_{\nu_{F2}}(t_{PS})}^2
\delta_{x_{BS1}+(t_{PS}-t_{BS1}),x_{PS}}
$
in the following form:
\begin{equation}
\Ket{\psi_{PS}} = 
\frac{1}{\sqrt{2}} 
\left(e^{i\phi_b} \Ket{t_{PS}+1,x_{PS},b} + 
e^{i\phi_b} \Ket{t_{PS}+1,x_{PS},a}\right).
\label{StatePS} % \eqno(40)
\end{equation}
Similarly, the state of the particle in the second beam splitter BS2, 
after a free evolution F3, with the conditional probability
$
\Mod{\chi_{\nu_{F3}}(t_{PS})}^2
\Mod{\chi_{\nu_{F3}}(t_{BS2})}^2
\delta_{x_{PS}+(t_{BS2}-t_{PS}),x_{BS2}}
$
is represented by
\begin{equation}
\Ket{\psi_{BS2}} = 
-\frac{1}{2}(e^{i \phi_b}+e^{i \phi_a}) \Ket{t_{BS2},x_{BS2}+1,a} 
-\frac{1}{2}(e^{i \phi_b}-e^{i \phi_a}) \Ket{t_{BS2},x_{BS2}+1,b}. 
\label{StateBS2} % \eqno(41)
\end{equation}
Finally, according to (\ref{FinalState}), the state in the detector D after
a free evolution F4 is:
\begin{equation}
\Ket{\psi_{D}} = \Ket{t_D,x_{D_{\sigma_D}},\sigma_D}.
\label{StateD} % \eqno(42)
\end{equation}
The corresponding conditional probabilities of finding the particle in the 
output channels $a$ and $b$ are:
\begin{equation}
\Mod{\chi_{\nu_{F4}}(t_{BS2})}^2
\Mod{\chi_{\nu_{F4}}(t_D)}^2
\delta_{x_{BS2}+(t_D-t_{BS2}),x_D}\
\delta_{\sigma_D,a}\; \cos^2\frac{\phi_a-\phi_b}{2} 
\label{ProbDa} % \eqno(43)
\end{equation}
and
\begin{equation}
\Mod{\chi_{\nu_{F4}}(t_{BS2})}^2
\Mod{\chi_{\nu_{F4}}(t_D)}^2
\delta_{x_{BS2}+(t_D-t_{BS2}),x_D}\
\delta_{\sigma_D,b}\; \sin^2\frac{\phi_a - \phi_b}{2},
\label{ProbDb} % \eqno(44) 
\end{equation}
respectively.

Observation:  if  there were no phase shifter on the evolution path, 
the second detector would never detect the particle, which is in 
agreement with the experimental observation.

Making use of the expression (\ref{ProbDet}) one can write the total,
conditional probability of detecting the particle at $\tau_D$, in either $a$ or
$b$ channel:  
\begin{eqnarray}
&&\Prob(\tau_D;\nu_Z,\nu_{F1},\nu_{BS1},\nu_{F2},\nu_{PS},\nu_{F3},
\nu_{BS2},\nu_{F4},\nu_D) = 
\left|\BraKet{t_Z,x_Z,a}{\psi_0}\right|^2\nonumber \\
&&\quad 
\Mod{\chi_{\nu_{F1}}(t_Z)}^2
\Mod{\chi_{\nu_{F1}}(t_{BS1})}^2 
\Mod{\chi_{\nu_{F2}}(t_{BS1})}^2
\Mod{\chi_{\nu_{F2}}(t_{PS})}^2
\Mod{\chi_{\nu_{F3}}(t_{PS})}^2
\Mod{\chi_{\nu_{F3}}(t_{BS2})}^2
\Mod{\chi_{\nu_{F4}}(t_{BS2})}^2
\Mod{\chi_{\nu_{F4}}(t_D)}^2 \nonumber \\
&&\quad \delta_{x_Z+(t_{BS1}-t_Z),x_{BS1}}
\delta_{x_{BS1}+(t_{PS}-t_{BS1}),x_{PS}}
\delta_{x_{PS}+(t_{BS2}-t_{PS}),x_{BS2}} \nonumber\\
&&\quad \left[\delta_{x_{BS2}+(t_D-t_{BS2}),x_{Da}}
            \delta_{\sigma_D,a}\; cos^2\frac{\phi_a - \phi_b}{2}
+           \delta_{x_{BS2}+(t_D-t_{BS2}),x_{Db}}
            \delta_{\sigma_D,b}\; \sin^2\frac{\phi_a - \phi_b}{2} \right].
\label{ProbDfull} % \eqno(46)  
\end{eqnarray}
The probability (\ref{ProbDfull}) can be nonzero only for  $t_D=t_Z+(x_D-x_Z)$.
It implies that the final state  (\ref{StateD}) is chosen by the Nature, with
the probability (\ref{ProbDfull}), as one of two vectors:
\begin{equation}
\Ket{\psi_D;\sigma_D}=
\begin{cases}
\Ket{t_D=t_Z+(x_D-x_Z),x_{D_{a}},a} & \text{for}\ \sigma_D=a, \\
\Ket{t_D=t_Z+(x_D-x_Z),x_{D_{b}},b} & \text{for}\ \sigma_D=b  
\end{cases}
\label{StateDsigma} % \eqno(47)
\end{equation}

Because the physical time is here a dynamical variable one can construct
different observables characterizing time behavior of our system. One of the
most interesting is the observable which measure a possibility of finding the
particle at given time in the channel $\sigma$. The corresponding decomposition
of unity can be written as:
\begin{equation}
M_T(t,\sigma)=\sum_x\, \Ket{t,x,\sigma}\Bra{t,x,\sigma},
\label{MTSigma} % \eqno(48)
\end{equation}
where $t \in \Rnumb$ and $\sigma=a,b$. The conditional probability of finding 
the particle at given time in the channel $\sigma$, while the  particle is in 
a state $\rho$, can be calculated according to the usual formula:
\begin{equation}
\Prob(\rho;t,\sigma)= \Trace(M_T(t,\sigma)\rho).
\label{Prob_rho_sigma} % \eqno(49)
\end{equation}
In our, very simple model the conditional probability (\ref{Prob_rho_sigma})
calculated  for the state (\ref{StateDsigma}) is non--zero only if $t=t_D$ and
$\sigma=\sigma_D$. In this case the probability is equal to one for
$t_D=t_Z+(x_D-x_Z)$. It means that  $t_D$ can be interpreted as time of
detection. In more realistic model the final state can have quite complicated
distribution in time.

Full probability of detection of the particle which passed the evolution path
$h$ (shown above the equation (\ref{FinalState})) is the product of all the
conditional probabilities along the  evolution path  and the conditional
probability given by (\ref{Prob_rho_sigma}). In our case this procedure gives,
in fact, the probability  (\ref{ProbDfull}).

In this toy model the states of the particle are localized in time by all
devices. The only exception is free evolution which gives states spread over
the time. 

All the moments which appear in this evolution process like $t_Z$, which 
represents the time of particle emission from the source,  $t_{BS1}, t_{PS},
t_{BS2}$ and $t_D$ which denote the times of arrival to  the appropriate
devices, are random variables. In addition, their probability  distributions
are not independent  e.g., the probability of choosing by the
"World Lottery" the time of arrival to the first beam splitter $t_{BS1}$ is
dependent on time of emission from the source $t_Z$, and so on. 
These are special features of the model considered above.

As an example let us calculate the average  time at which the particle can be
localized in $x_{BS1}$ (the first beam splitter) under the condition that the
particle was emitted from the source at the fixed time $t_Z$.

In general, for any evolution parameter $\tau_F$, where $0<\tau_F \leq
\tau_{BS1}$, the state can be obtained by two-step evolution according to  the
evolution operator (\ref{PEvTotal}). It can be written as:
\begin{equation}
\rho(\tau_F;t_Z,\nu_F)
\sum_{t_1,t_2}\chi_{\nu_F}(t_1)\chi_{\nu_F}(t_2)^\star
\Ket{t_1,x_Z-t_Z,a}\Bra{t_2,x_Z-t_Z,a}.
\label{State_Source_BS1} % \eqno(50)
\end{equation}
To calculate the average time of finding the particle in the place where is the
located the first beam splitter, we have to define the appropriate operators. 
First, we define the decomposition of unity which is able to determine the
position of particle in the space--time but independently of channels:
\begin{equation}
M_G(t,x)=\sum_{\sigma} \Ket{tx\sigma}\Bra{tx\sigma},
\label{MGOp} % \eqno(51)
\end{equation}
The observable which measure the time at fixed coordinate point $x$ can be 
thus defined as:
\begin{equation}
\mathcal{T}(x)=\sum_{t} t M_G(t,x).
\label{Time_xOp} % \eqno(52)
\end{equation}
The average time of finding the particle at $x_{BS1}$ can be  calculated
according to the usual rules as:
\begin{eqnarray}
\langle\mathcal{T}(x)\rangle &=& 
\Trace(\mathcal{T}(x)\rho(\tau_F;t_Z,\nu_F)) \nonumber\\
&=& \sum_{t'} t' |\chi_{\nu_F}(t')|^2 \delta_{x_Z-t_Z+t',x_{BS1}} \nonumber\\
&=& (x_{BS1}-x_Z+t_Z) |\chi_{\nu_F}(x_{BS1}-x_Z+t_Z)|^2.
\label{AverTimeArrival} % \eqno(53)
\end{eqnarray}
This time can be also called the time of arrival because the particle is
emitted at the fixed time $t_Z$ and the difference
$\langle\mathcal{T}(x)\rangle - t_Z$ should give the average time of fly
between the source and the first beam splitter.

We obtain the expected result, though with the additional factor
$|\chi_{\nu_F}(x_{BS1}-x_Z+t_Z)|^2$. The time of arrival is proportional to
distance  between the source and the beam splitter and it is what we expected,
because in this very simple model there is no spreading of "wave packets" over
the space.  The additional coefficient is equal to the probability of finding
the particle at a given point of   time  axis. In the case of Schr\"odinger
equation, as an equation of motion (implemented by the projection operators
like those for free motion, see (\ref{PhiBase},\ref{PEvFree})), the functions 
$|\chi_{\nu}(t)|=1$.

Taking into account that times of emissions $t_Z$ of particles, in principle,
are  also random variables, the probability that a particle being in the initial
state $\Ket{\psi_0}$, can be localized at a given time  in $x_{BS1}$ is the
product of $|\BraKet{t_Z,x_Z,a}{\psi_0}|^2$ and 
$\Trace(M_G(t,x_{BS1})\rho(\tau_F;t_Z,\nu_F))$. This allows to derive  the
average time $<t_{ar}>$ of finding the particle at $x_{BS1}$  averaged, in
addition, over the emission times $t_Z$ as:
\begin{eqnarray}
<t_{ar}> &=& 
\sum_{t,t_Z}\, t\, 
|\BraKet{t_Z,x_Z,a}{\psi_0}|^2 \, \Trace(M_G(t,x_{BS1})\rho(\tau_F;t_Z,\nu_F)) 
\nonumber\\
&=& \sum_{t_Z} |\BraKet{t_Z,x_Z,a}{\psi_0}|^2\,  
|\chi_{\nu_F}(x_{BS1}-x_Z+t_Z)|^2 \, (x_{BS1}-x_Z+t_Z).
\label{AverAverTimeArrival} % \eqno(54)
\end{eqnarray}

\vspace{1cm}

\leftline{Toy Model II: Beam splitter of unitary--type}
\medskip 
As the second model we consider the same one--dimensional Mach--Zehnder
interferometer as in the Fig.~\ref{fig2}. We assume the same space of states as
in the Model I. 

The main difference between both models is that in the second one we treat both
beam splitters and phase shifter as a single device. It is described by the
evolution operator similar to the description of free motion in the first
model.

The source of the particles we describe exactly  by the same decomposition of
unity as in the Model I, see (\ref{PEvZ}). 

On the other hand, both beam splitters and the phase shifter we describe by the
operator:
\begin{equation}
\Eop_{DEV}(\nu) = 
\sum_{t,x,\zeta} \Ket{\phi_{\nu,x,\sigma}(t',x',\zeta)}
                 \Bra{\phi_{\nu,x,\sigma}(t',x',\zeta)},
\quad \text{where } \nu \in \Znumb,
\label{PEvDev} % \eqno(55)		    
\end{equation}
where $\phi_{\nu,x,\sigma}$ is given by the equation (\ref{PhiBase}), but with
another operator $\hat{S}(t)=\hat{S}^t$. The operator $\hat{S}$ describes all
the three devices at once, see \cite{Grif03}:
\begin{equation}
\hat{S}\Ket{x,\sigma} = 
\begin{cases}
\frac{1}{\sqrt{2}}(\Ket{x_{BSk}+1,b}+\Ket{x_{BSk}+1,a}) & \text{for}\  
x=x_{BSk},\ \sigma=a \text{ and } k=1,2, \\
\frac{1}{\sqrt{2}}(-\Ket{x_{BSk}+1,b}+\Ket{x_{BSk}+1,a}) & \text{for}\  
x=x_{BSk},\ \sigma=b \text{ and } k=1,2, \\
e^{i\phi_\sigma}\Ket{x_{PS}+1,\sigma}& \text{for}\  
x=x_{PS},\ \sigma=a,b \text{ and }, \\
\Ket{x+1,\sigma} & \text{otherwise}. 
\end{cases}
\label{SDev} % \eqno(56)
\end{equation}
It means that a single drawing of lots by the Nature decides about the 
evolution through all three devices. In this case, there are no intermediate
times like   $t_{BS1}, t_{PS}$ and  $t_{BS2}$ which are effects of subsequent
steps of the evolution.

The last step of the evolution is again determined by the detectors and it is
described  by the operator (\ref{PEvD}).

In this case the projection evolution operator is shorter and can be written as
\begin{equation}
\Mew{\tau}{\nu} = 
\begin{cases}
\Eop_Z(\nu) & \text{for}\    \tau=0, \\
\Eop_{DEV}(\nu)      & \text{for}\  0 < \tau < \tau_{D} \\
\Eop_D(\nu) & \text{for}\    \tau=\tau_{D}.
\end{cases}
\label{PEvMod2}  % \eqno(57)
\end{equation}

In this model there are  only  3 steps of the evolution. The system starts from
the emission of the particle, passes all the three devices  (the first beam
splitter,  the phase shifter and the second beam splitter)  and  ends at  the
detectors: 
\begin{equation}
\rho(\tau_D;\nu_Z,\nu_{DEV},\nu_D) = 
\ \frac{
\Eop_D(\nu_D)\Eop_{DEV}(\nu_{DEV})\Eop_Z(\nu_Z)
\rho_0
\Eop_Z(\nu_Z)\Eop_{DEV}(\nu_{DEV})\Eop_D(\nu_D)}{
\Trace\left[  
\Eop_D(\nu_D)\Eop_{DEV}(\nu_{DEV})\Eop_Z(\nu_Z)
\rho_0
\Eop_Z(\nu_Z)\Eop_{DEV}(\nu_{DEV})\Eop_D(\nu_D)
\right]}.
\label{StateRho2}  % \eqno(58) 
\end{equation}
The probability distribution of finding the particle
at the time $t_D$,  in a given output channel,
assuming the same initial state (\ref{InitState}),
when it follows a given evolution path 
$h=(\nu_Z=t_Z,\nu_{DEV}, \nu_D=(t_D,\sigma_D))$
is
\begin{eqnarray}
\Prob(\tau_D;\nu_Z,\nu_{DEV},\nu_D) = 
&&\Trace\left[  
\Eop_D(\nu_D)\Eop_{DEV}(\nu_{DEV})\Eop_Z(\nu_Z)
\rho_0
\Eop_Z(\nu_Z)\Eop_{DEV}(\nu_{DEV})\Eop_D(\nu_D)
\right] = \nonumber \\ 
&&
\left|\BraKet{t_Z,x_Z,a}{\psi_0}\right|^2
\Mod{\chi_{\nu_{DEV}}(t_Z)}^2
\Mod{\chi_{\nu_{DEV}}(t_D)}^2 
\left|\Bra{x_{D},\sigma_D}S^{\Delta t}\Ket{x_Z,a}\right|^2
\label{ProbDet2}  % \eqno{59}
\end{eqnarray}
where $\Delta t = t_D-t_Z$ denotes the difference between the time of emission
of the particle and the time of its detection.

Let us explicitly write the action of the unitary operator $S^{\Delta t}$ onto
the state $\Ket{x_Z\,a}$:
\begin{equation}
S^{\Delta t}\Ket{x_Z\,a}=
\begin{cases}
\Ket{x_Z+\Delta t,a} 
	&\text{for}\ 0\leq \Delta t < x_{BS1}-x_Z\\
\frac{1}{\sqrt{2}} (\Ket{x_Z+\Delta t,b}+\Ket{x_Z+\Delta t,a})
	&\text{for}\ x_{BS1}-x_Z \leq \Delta t < x_{PS}-x_Z\\	
\frac{1}{\sqrt{2}} (e^{i \Phi_b}\Ket{x_Z+\Delta t,b}+
					e^{i \Phi_a}\Ket{x_Z+\Delta t,a})
	&\text{for}\ x_{PS}-x_Z \leq \Delta t < x_{BS2}-x_Z\\
\frac{1}{2}\{(-e^{i \Phi_b}+e^{i \Phi_a})\Ket{x_Z+\Delta t,b} +
(e^{i \Phi_b}+e^{i \Phi_a})\Ket{x_Z+\Delta t,a}\}	
	&\text{for}\  \Delta t \geq x_{BS2}-x_Z\\
\end{cases}
\label{SActionS}   % \eqno(60)
\end{equation} 
Using above expression, the total conditional probability (\ref{ProbDet2})  can
be easily rewritten as:
\begin{eqnarray}
\Prob&&(\tau_D;\nu_Z,\nu_{DEV},\nu_D) = 
\left|\BraKet{t_Z,x_Z,a}{\psi_0}\right|^2
\Mod{\chi_{\nu_{DEV}}(t_Z)}^2
\Mod{\chi_{\nu_{DEV}}(t_D)}^2 
 \nonumber \\ 
&&
\quad 
\left[\delta_{x_Z+(t_D-t_Z),x_{Da}}
      \delta_{\sigma_D,a}\; cos^2\frac{\phi_a - \phi_b}{2}
+      \delta_{x_Z+(t_D-t_Z),x_{Db}}
       \delta_{\sigma_D,b}\; \sin^2\frac{\phi_a - \phi_b}{2} \right].
\label{ProbDfull2} % \eqno(61)  
\end{eqnarray}
The  result  is very similar to those from the Model I, see (\ref{ProbDfull}),
especially for the case when the functions $\chi_\nu(t) \approx 1$. As in the
previous case,  the interval between  the emission of the particle by the
source and its  detection is determined by distance between the source and the
detector.  The main difference between both models  is in timing  of
evolution  steps. In the  Model II, besides of $t_Z$ and $t_D$ there are no
intermediate times (which are random variables) and  additional factors in
probabilities  related to them.  Lack of spreading in time  is determined
here by the "toy" form of the operator $S$ (\ref{SDev}). 

In this notes we have compared only two models of the 
Mach-Zehnder interferometer within the projection evolution approach. In
practice, they give nearly the same results of the evolution. However, it seems
that up to date experiments show that, the beam splitters and phase shifters
are closer to the second model than to the first one.  In both models the
physical time is a dynamical variable which is changing in rather primitive
(nearly no smearing in time) way. Because of this, they cannot show all
features of dynamics of  time which is in principle a random variable in the
evolution process.

\section{Summary}
In this paper, we propose  a fundamental mechanism of quantum evolution based
on the idea of  "natural measurements" performed by the Nature at each step
of changes (evolution) of the physical systems. 

This scheme leads to unification of unitary evolution and measurements
handled by the so--called projection postulate. 

 As a special case, our postulate allows to reproduce the Schr\"odinger
equation and go beyond it. In this way one can reproduce not only the
Schr\"odinger equation but also other quantum equations of motions e.g., the
relativistic equations.

Obviously, the \PEv postulates require  many tests, yet. 
However, it seems that for each physical system one can find an appropriate set
of projection operators which allows to apply the \PEv postulate to get 
physical states. A possible method to construct the \PEv operators is the
method of generating operators. 

It is also important to notice that the \PEv postulate gives 
the unique  opportunity to treat space and time on equal footing. 
This feature can open  some new directions of development of quantum theory.

In the preprint we left many open questions like the problem of time operator,
the time--energy uncertainty relation, problem of interactions (potentials) in
time variable and many others.

\end{document}